\documentclass[10pt]{article}

\def\ref#1{\cite{#1}}

\def\giorno{}

\def \grad {\nabla}
\def \ov{\over}
\def \pd{\partial}
\def \pn{\par\noindent}
\def \bs{\bigskip}

\def\.#1{\dot #1}

\def\L{{\cal L}}
\def\cR{{\cal P}}
\def\R{{\bf R}}  
\def\S{{\cal S}}

\def\ss{\subset}
\def \pa{\partial}
\def \pd{\partial}
\def\=#1{\bar #1}
\def\~#1{\widetilde #1}
\def\.#1{\dot #1}
\def\^#1{\widehat #1}

\def\d{{\rm d}}       

\def\grad{\nabla}     
\def\({\left(}
\def\){\right)}
\def\[{\left[}
\def\]{\right]}

\def \tr {transformation}
\def \sy {symmetry}
\def \sys {symmetries}

\def\a{\alpha}
\def\b{\beta}

\def\phi{\varphi}
\def\la{\lambda}

\def\phi{\varphi}

\def\De{\Delta}
\def\th{\lambda}

\def\car{{${\rm Cari\~nena}$}}
\def \qq{\qquad}
\def \q{\quad}

\def \Z{Zhdanov}
\def\ba{B\"acklund}

\def\so {solution}
\def \id{\! :=}
\def \df{differential }
\def \eq{equation}

\begin{document}

\title{\bf {Partial Lie-point symmetries \\ of differential equations \\ ~ }}

\author{Giampaolo Cicogna \\ 
{\it Dipartimento di Fisica, Universit\`a di Pisa} \\ 
{\it Via Buonarroti 2, Ed.B, 56127 Pisa (Italy)} \\
{\tt cicogna@df.unipi.it} \\ ~ \\
Giuseppe Gaeta \\ 
{\it Dipartimento di Fisica, Universit\`a di Roma} \\
{\it P.le A. Moro 2, 00185 Roma (Italy)} \\
{\tt gaeta@roma1.infn.it} }

\date\giorno

\maketitle


\begin{abstract}
When we consider a differential equation $\Delta=0$ whose set of
solutions is $\S_\Delta$, a Lie-point exact symmetry of this is a Lie-point
invertible transformation $T$ such that $T(\S_\Delta)=\S_\Delta$, i.e. such that any
solution to $\Delta=0$ is tranformed into a (generally, different) solution to
the same equation; here we define {\it partial} symmetries of $\Delta=0$ as 
Lie-point invertible transformations $T$ such that there
is a nonempty subset $\cR \subset \S_\Delta$ such that $T(\cR) = \cR$, i.e.
such that there is a subset of solutions to $\Delta=0$ which are transformed
one into the other. We discuss how to determine both partial symmetries
and the
invariant set $\cR \subset \S_\Delta$, and show that our procedure is effective
by means of concrete examples. We also discuss relations with conditional
symmetries, and how our discussion applies to the special case of dynamical
systems. Our discussion will focus on continuous Lie-point partial
symmetries, but our approach would also be suitable for more general 
classes of transformations; the discussion is indeed extended to partial 
generalized (or Lie-B\"acklund) symmetries along the same lines, and 
in the appendix we will discuss the case of discrete partial symmetries.
\end{abstract}

\vfill\eject

\section*{Introduction}

Symmetries are most useful in the study of differential equations
\ref{BK,Gae,Olv,Ste,Win1,Win2,Win5}, and they are useful in different
ways.

On the one hand, one can consider symmetry reduction of differential
equations and thus obtain classes of exact solutions; on the other
hand, by definition, a symmetry of a differential equation transforms
solutions into solutions, and thus symmetries can be used to generate
new solutions from  known ones.

We remark that these transformations can yield highly nontrivial solutions
starting from very simple known ones; the best example is maybe provided
by the symmetry transformation taking the trivial constant solution of the
heat equation into the fundamental (gaussian) one, as discussed in section
2.4 of \ref{Olv}.

Transformations which act point-like in the space of independent and
dependent variables are also called Lie-point symmetries, to be
distinguished from transformations of more general nature,
involving e.g. derivatives or integrals of these variables \ref{Olv}.

Several generalizations of standard Lie-point symmetries have been
considered in the literature; we mention in particular the {\it
conditional
symmetries}, first introduced by Bluman and Cole \ref{BlC1,BlC2} and then
systematized and applied by Levi and Winternitz 
\ref{LW,Win3,Win4,Win5}, i.e. transformations $T$ which are {\it not }
symmetries of the given differential equation, but such that this has
some solution $u_0$ invariant under them,
$T [u_0] = u_0$. The trivial example of this phenomenon is provided by
non-symmetric equations admitting the zero solution: this has obviously
a very large group of symmetries.

Much less trivial -- and actually very useful -- examples are considered
in the literature; we refer e.g. to \ref{Win5} for both a general
discussion of the method and for relevant examples; this reference also
lists a
number of works where conditional symmetries have been applied to
equations of physical relevance.

It should be noted that, in a way, conditional symmetries focus on (an
extension of) the first use of symmetries mentioned above, i.e. on the
search for invariant solutions.

One should also mention that conditional symmetries are strictly related
to the so called ``direct method" of Clarkson and Kruskal \ref{ClK}; the
group
theoretical understanding of this \ref{LW} involves conditional
symmetries and is related to the ``non-classical method" of Bluman and Cole
\ref{BlC1,BlC2} and to the ``side conditions" of Olver and Rosenau
\ref{OlR}. For the relation between conditional symmetries and the
other mentioned approaches, see also the discussion by Pucci and
Saccomandi \ref{PuS} (and their recent paper \ref{PuS3}). For a related 
approach, see also the work
of Fushchich and collaborators \ref{Fus}. In \ref{Cra1}, a symmetry
formulation was used to transform PDEs on a bounded domain of $\R^n$ with
boundary conditions into PDEs on a boundaryless manifold, the role of
boundary conditions being now played by an extra symmetry condition; this 
approach has been applied e.g. in \ref{Cra2}.

In this paper, we propose an extension of the notion of symmetries which
goes in the direction of the other use, i.e. of transformations taking
solutions to solutions. More specifically, let  $\S_\De$ be the 
set of all solutions to a given differential equation $\Delta=0$; an exact 
symmetry will be a transformation $T$ which induce an action (in the 
suitable function space \footnote{A more geometrical definition is also
possible: the transformation, acting in the space of independent and
dependent variables, has to map the graphs of solutions to graphs of,
generally different, solutions; see \ref{Olv}.}) transforming each element 
$s \in \S_\Delta$ into a -- generally, different -- element 
$s'\in \S_\Delta$.

We will consider {\it partial symmetries}: these will be transformations
$T$ which induce an action (in the suitable function space) transforming
each element $u \in \cR \subset \S_\De$ into a -- generally, different --
element $u' \in \cR \subset \S_\De$, for $\cR$ some non-empty subset of
$\S_\Delta$  (a more precise definition will be given below). Notice that
when $\cR = \S_\Delta$ we actually have standard exact symmetries, while 
when $\cR$ reduces to a single solution, or to a set of solutions each of 
them invariant under $T$, we
recover the setting of conditional symmetries. Indeed, as we discuss in
section 2, conditional symmetries will always be a special class of
partial symmetries.

The reader could like to see immediately a simple example in order to
better grasp this qualitative definition; here is one in terms of an
infinitesimal symmetry generator, for a PDE for $u = u(x,y)$.
\bigskip

{\it Example.}
Consider the vector field $X=(\pd / {\pd x})$,
which generates the translations of the variable $x$, and is an
exact \sy\ for any equation not depending explicitly on $x$. An
equation, for instance, like $x(u_x-yu_y)+u_{xy}-u_y+yu^2_{yy}=0$
certainly does not admit translational \sys\ $\pd/\pd x$ or $\pd/\pd y$;
however, it admits the particular family of \so s $u=y\exp(x+\th) $
which are transformed one into the other under the translations of the
variable $x$,
as this can be reabsorbed by a corresponding shift in the parameter $\th$.
Notice none of these solutions is invariant under the $x$ translation. 
$\odot$

\bs

It should be mentioned that this notion of ``partial symmetries" 
had been considered in general terms by several authors;
however, to the best of our knowledge, this has never gone beyond the
stage of introducing an abstract notion, without discussion of methods
to implement this -- nor {\it a fortiori } concrete applications --
except for conditional symmetries, i.e. invariant solutions.

Indeed, the notion of partial symmetries also appears in early general
definitions of conditional symmetries (the method being actually
implemented, however, only in the stricter sense mentioned above, the
notion of conditional symmetries passed to be used in the present
sense). We refer for this to the works by Vorob'ev \ref{Vor2,Vor3},
which also suggests a solution to some priority
question by mentioning this notion had already be considered by Felix
Klein \ref{Vor3}.

In the following we will characterize partial symmetries in an
operational way, and show -- in theoretical terms but also by means of
concrete examples -- how they can be used to obtain solutions to nonlinear
differential equations. We will mainly focus, as customary in symmetry
study of differential equations, on continuous transformations, and
actually study their infinitesimal generators (the reason being, as usual
in this field, that the equations we obtain for these are much more
palatable than those obtained by considering finite transformations);
discrete transformations will be briefly considered in the appendix.

It should be stressed that, as it also happens for conditional symmetries
(but contrary to the case of exact Lie-point symmetries), the
determining equations for (infinitesimal generators of) partial symmetries
will be nonlinear; we will thus be in general unable to determine all
the partial symmetries to a given equation;
thus in general we will need to have some
hint -- maybe on the basis of physical considerations -- of what the
partial symmetries could be, for the method to be applicable with
reasonable effort. Nevertheless, determination of one or some partial
symmetries can already be of use in the search for exact solutions
(as it also happens for exact symmetries).

Finally, we note that we focus our discussion on Lie-point symmetries, 
but the definition 
and discussion can be extended to generalized symmetries (sometimes also 
called Lie-\ba\ symmetries). We will also compare partial generalized 
symmetries with conditional generalized symmetries (for these, see 
\ref{Zh}).

To conclude this introduction, we stress that the partial symmetry method can 
lead to a class of solutions including solutions which cannot be obtained 
either by exact symmetries or by conditional symmetries. The examples below 
show that this is actually the case. 

\bs

The {\it plan of the paper} is as follows. In section 1 we introduce and
define partial symmetries in a constructive way; i.e., identifying also
the subset $\cR$ of solutions which is left globally invariant by the
partial symmetry and explaining how to compute this and the symmetry
itself.
In section 2 we discuss the relation between these partial symmetries
and the conditional symmetries of Levi and Winternitz, and with the 
conditional generalized symmetries of Zhdanov. In section 3 we
discuss how, in certain circumstances, given partial symmetries guarantees
the considered differential equation enjoys a ``partial superposition
principle" (defined there). In section 4 we specialize our discussion to
the case of dynamical systems. The final sections are devoted to a detailed
discussion of concrete  examples: these deal with PDEs of interest for 
Physics in section 5, and with dynamical systems in section 6. 
As already mentioned, our discussion 
is at the level of
infinitesimal symmetry generators and thus continuous partial symmetries,
but in the appendix we briefly discuss discrete partial symmetries.

\bs\pn
{\bf Acknowledgements.} We would like to warmly thank Decio Levi for some
useful discussions, right suggestions (in particular, 
remark 3 is due to him), and encouragement. We also thank
prof. P.J. Olver for a useful (E-mailed) comment, and prof. J.F. \car\ 
for sending us the preprint version of his forthcoming book.
Finally, we thank unknown referees for pointing out the relations
of our work to a geometrical approach by Gardner (see remark 5 below),
and for encouraging us to explicitely consider generalized symmetries as well.

\section{Partial symmetries of differential problems}
\def\en#1{\eqno(1.{#1})}

\bs
Let us consider a general differential problem, given in the form of a
system of $\ell$ \df \eq s, and shortly denoted, as usual, by
$$\De\id\De(x,u^{(m)})\ =\ 0\en{1}$$
where $\De\id(\De_1,\De_2,\ldots,\De_\ell)$ are smooth functions involving
$p$ independent variables $x\id(x_1,\ldots,x_p)\in\R^p$ and $q$
dependent ``unknown'' variables $u\id(u_1,\ldots,u_q)\in\R^q$,
together with the derivatives of the $u_\a$ with respect to the $x_i$
($\a=1,\ldots,q;\ i=1,\ldots,p$) up to some order $m$.

Let
$$X=\xi_i{\pd\ov{\pd x_i}}+\phi_\a{\pd\ov{\pd u_\a}}
\qq\qq \xi_i=\xi_i(x,u),\q\phi_\a=\phi_\a(x,u)\en{2}$$
be a given vector field, where $\xi_i$ and $\phi_\a$ are $p+q$ smooth
functions. For notational simplicity, we will shortly
denote by $X^*$ the ``suitable'' prolongation of
$X$, i.e. the prolongation which is needed when one has to consider its
application to the \df problem in consideration.
Alternatively, we may consider $X^*$ as the infinite prolongation of $X$,
it is clear indeed that only a finite number of terms are required
and will appear in all the actual computations. As well known \ref{Olv}, 
the vector field $X$ is (the Lie generator of) an  exact \sy\ of the \df
problem (1.1) if and only if \footnote{We will always assume that the 
standard technical condition of ``maximal rank'' is satisfied \ref{Olv};
note that this has been recently relaxed \ref{AFT}.}
$$X^*\ \De\Big|_{\De=0}\ =\ 0 \en{3}$$
i.e. if and only if the prolongation $X^*$ (here obviously,
$X^*={\tt pr}^{(m)}(X)$, the $m-$th prolongation of $X$) applied to the \df
operator $\De$ defined by (1.1) vanishes once restricted to the set
$S^{(0)}\id\S_\Delta$ of the \so s to the problem $\De=0$.

We now assume that the vector field $X$ is {\it not} a \sy\ of (1.1), hence
$X^*\De\Big|_{S^{(0)}}\not=0$: let us put
$$\De^{(1)}\id X^*\De\ .\en{4}$$
This  defines  a \df operator $\De^{(1)}$, of order $m'$ not greater
than the order $m$ of the initial operator $\De$. Assume now that the set
of
the simultaneous solutions of the two problems
$\De=0$ and $\De^{(1)}=0$
is not empty, and let us denote by $S^{(1)}$ the set of these solutions.
It can happen (see Examples 2--5) that this set is mapped into itself
by the \tr s  generated by $X$: {\it this situation is characterized
precisely by the property}
$$X^*\De^{(1)}\Big|_{S^{(1)}}\ =\ 0\en{5}$$
Then, in this case, we can conclude that,  although $X$ is not a \sy\ for
the full problem (1.1), it generates anyway a \tr\ which leaves globally
invariant a family of \so s of (1.1): this family is precisely $S^{(1)}$.

But it can also happen that $X^*\De^{(1)}|_{S^{(1)}}\not=0$ (see Examples 1
and 2), we then put
$$\De^{(2)}\id X^*\De^{(1)}\en{6}$$
and look for the \so s of the system
$$\De=\De^{(1)}=\De^{(2)}=0\en{7}$$
and repeat the argument as before: if the set $S^{(2)}$ of the \so s of
this system is not empty and satisfies in addition the condition
$$X^*\De^{(2)}\Big|_{S^{(2)}}\ =\ 0\en{8}$$
then $X$ is a \sy\ for the  subset $S^{(2)}$ of \so s of the initial
problem (1.1), exactly as before.

Clearly, the procedure can be iterated, and we can say:
\bigskip
\pn
{\bf Proposition 1}. {\it Given the general \df problem (1.1) and a
vector field (1.2), define, with
$\De^{(0)}\id\De$,
$$\De^{(r+1)}\id X^*\De^{(r)} \ . \en{9}$$
Denote by $S^{(r)}$ the set of the
simultaneous \so s of the system
$$\De^{(0)}=\De^{(1)}=\ldots=\De^{(r)}=0\en{10}$$
and assume this is not empty for $r \le s$.
Assume moreover that
$$\cases{
X^*\De^{(r)}\Big|_{S^{(r)}}\not= 0 &for $r=0,1,\ldots, s-1$ \cr
X^*\De^{(s)}\Big|_{S^{(s)}}\ =\ 0 & \ . \cr} \en{11}$$
Then the set $S^{(s)}$ provides a family of \so s to the initial problem
(1.1) which is mapped into itself by the \tr s generated by $X$.}

\bs
We shall say that $X$ is a ``partial \sy '', or $P$-symmetry for short,
for the problem~(1.1), and that the globally invariant subset of \so s
$\cR\id S^{(s)}$ obtained in this way is a ``$X-$symmetric set''.
We also refer to the number $s$ appearing in the statement as the order of 
the $P$-symmetry.

It is clear that, given a \df problem and a vector field $X$, it can
happen that the above procedure gives no result, i.e. that at some $p-$th
step the set $S^{(p)}$ turns out to be empty. Just to give an example,
consider,  with $\ell=1,\ q=1,\ p=2$ and putting
$x_1=x,\ x_2=~y$, the PDE
$$    x u_x + x^2 u_y + 1\ =\ 0\en{12}$$
and the vector field, generating the translations along the variable $x$,
$$X={\pd\ov{\pd x}}\en{13}$$
It is easy to verify that, after two steps of the above procedure, one
obtains an inconsistent condition. Then, in this case, we simply conclude
that (1.13) is not a $P-$symmetry for the problem (1.12).

Assume instead that a vector field $X$ is a $P-$symmetry for a given 
problem,
and therefore that a non-empty set $S^{(s)}$ of $X-$symmetric \so s
has been found.  We stress that the solution in this set  are, in general,
{\it not } $X$-invariant: only the set $S^{(s)}$ is globally invariant,
while the solutions are transformed one into the other under the $X$
action (see the simple example reported in the introduction, and the 
examples in sections 5 and 6). Note that when there is some solution $u_0$ 
which is invariant under
a given $P$-symmetry $X$, then $X$ is also a {\it conditional symmetry}
for the differential problem  at hand (see next section for a discussion
of this point).  The set of \so s in $S^{(s)}$ will be constituted by one
or more {\it orbits}
under the action of the one-parameter Lie group $\exp(\la X)$,
and (apart from the trivial case of the $X-$invariant \so s) each
one of these orbits can be naturally parametrized by
the real Lie parameter $\la$.
Denoting by  $u^{[\la]}\id u(x;\la)$ the \so s belonging to any
given orbit, one has that each family $u^{[\la]}$ satisfies the
differential equation (in ``evolutionary form'' \ref{Olv})
$$Q\ u^{[\la]}\ =\ {\d u^{[\la]}\ov{\d\la}}\ .\en{14}$$
where
$$Q=- \xi_i {\pd\ov{\pd x_i}}  + \phi_\a{\pd\ov{\pd u_a}}\ .\en{15}$$

\bs\pn
{\it Remark 1.} It should be recalled \ref{Olv} that when we
consider ordinary Lie-point
symmetries and pass from the infinitesimal (Lie algebra) level to the
finite (Lie group) one, in general we have only a {\it local} Lie group,
i.e. the map $T_\lambda = e^{\lambda X}$ is a symmetry of the
differential equation only for $\la$ in some interval
$|\lambda | < c_0$. Clearly, the same will apply here.
$\odot$

\bs\pn
{\it Remark 2.}
In the determination of $P$-symmetries and of orbits of solutions, the
``higher" equations $\Delta^{(r)}$ (with $r \not=0$) in the hierarchy
will only be considered on the submanifold $S^{(p)}$, $p\le r$; thus we can
also consider them directly on $S^{(r)}$ from their introduction (and
further
restrict them if the procedure has to go on). This has no conceptual
advantage, but can be appropriate for computational ease, as we will see
below in the examples.  $\odot$

\bs\pn
{\it Remark 3.}
Our above procedure can be given a nice geometrical interpretation;
to discuss this, we focus on finite action of the vector field $X$ (1.2),
or more precisely of its prolongation $X^*$, on the differential operator
(1.1). This is given by
$$ e^{\th X^*} \De = \De + \th X^* \De + {\th^2\ov{2!}}(X^*)^2 \De +
... \ = \
\sum_{k=0}^\infty {\th^k \over k !} (X^*)^k \De \ . \en{16} $$
where $\th$ is the Lie parameter. If $X$ is an exact symmetry of $\De$,
this must be zero whenever $\De =0$; indeed
we know that with $X$ a symmetry, $X^* \De =0$ on the solution set
$S^{(0)}$, and therefore a fortiori $(X^*)^k \De =0$ on this same set.

Now we notice that with our construction, $\De^{(0)}\id\De$,
$\De^{(1)}=X^* (\De)$, $\De^{(2)} = X^*
\big(X^* (\De)\big)$, and so on; our condition that $X^*$ is a symmetry for
the solution set $S^{(s)}$ is then
$$(X^*)^s\De\Big|_{S^{(s)}}=0\ .\en{17}$$
Looking back at (1.16), we rewrite it in the form
$$ e^{\th X^*} \De^{(0)} \ = \ \De^{(0)} \ + \ \sum_{r=1}^{s-1} \,
{\th^r\ov{r!}}\De^{(r)} \ + \
\sum_{k=s}^{\infty}\, {\th^k \over k!} (X^*)^k \De^{(0)} \ ; \en{18}$$
Now, the conditions (1.10) on the chain of equations $\De^{(r)}=0,\
r=0,1,\ldots,s-1$, together with (1.11) or (1.17) show that the r.h.s. of
(1.18) does actually vanish on the solution set $S^{(s)}$.
In other words, the fact that $X$ is a  partial symmetry  for
$\De^{(0)}=0$ guarantees that the second sum in (1.18)
vanishes; requiring by hand the vanishing of each term in the first sum
guarantees that the whole series vanishes, and thus identifies a set of
conditions sufficient to guarantee that each solution of $\De^{(0)}=0$
complying with these conditions is transformed into another -- in general
different -- solution of $\De^{(0)}=0$ complying with the same
conditions. $\odot$

\bs\pn
{\it Remark 4.} Our discussion is in terms of standard Lie-point
symmetries; it is well known that one can also introduce {\it strong}
symmetries \ref{Gae,Olv}, i.e. those for which $X^* (\De) = 0$ on all
the jet space (and not just on the solution manifold $\S_\De$, i.e. not
just when $\De =0$); the relation between standard and strong symmetries
has been clarified in \ref{CDW}. It should be quite clear that our
approach could also be reformulated in terms of strong symmetries and
``strong $P$-symmetries''; these would be defined by lifting the
restriction
to relevant solution manifolds and sets [e.g. in (1.5), (1.8) and (1.11)].
We will not discuss this setting, but note that in this case one would
still obtain a set of solutions to the original equation which is globally
$X$-invariant (we recall a strong symmetry is also a symmetry), but the
occurrence of such a set would be even rarer, as $X$ would be required to
be a strong symmetry of the system (1.10). On the other hand, computations
would be more straightforward, as one would not have to perform the
substitutions needed to implement restrictions to solution sets. $\odot$

\bs\pn
{\it Remark 5.} After the completion of this paper, one of the referees 
pointed out that our approach is related to the geometrical method of 
Gardner \ref{GK,HTT}, which is itself related to Cartan's ideas, and in 
particular with the concept of the $k-$stable vector fields. The 
relation between partial symmetries and Gardner's approach appears to be 
not trivial. We do not discuss this relationship here, but just mention 
that our focus is on symmetry properties, while Gardner's one is 
essentially on geometrical structures. $\odot$

\section{Partial symmetries and conditional \\ symmetries}
\def\en#1{\eqno(2.{#1})}

It should be noted that the procedure presented here is related to, but
quite different in spirit from, the standard {\it conditional symmetries}
approach in several ways.

Let us briefly recall what conditional symmetries   are and how they
are determined (see \ref{LW,Win3,Win4,Win5} for a more complete 
discussion). 
Given a $m$-th order differential equation
$\De (x,u^{(m)})=0$, we say that
$$ X \ = \ \xi_i {\pa \over \pa x_i} \ + \ \phi_\a {\pa \over \pa u_\a}
    \qq ( i=1,..,p ; \ \a = 1,...,q) \en{1} $$
is a conditional symmetry for $\De$ if there is some solution $u(x)$ to
$\De = 0 $ which is $X$-invariant. The $X$-invariance condition can be
written as
$$ \phi_\a (x,u) \ - \ \sum_{i=1}^p \, \xi_i (x,u) \, { \pa u_\a \over
   \pa x_i}\ =\ 0 \qq (\a = 1 , ... , q) \en{2} $$
so that $X$-invariant solutions to $\De=0$ are also solutions to the system
$$ \cases{\De (x,u^{(m)} ) \ = \ 0 & \cr \phi_\a (x,u) - \xi_i (x,u)
\, (\pd u_\a / {\pd x_i}) \ = \ 0 & \q ($\a = 1,...,q)$ \ .  } \en{3}$$

By construction, $X$ will be an ordinary Lie-point symmetry for the
system (2.3). Note that $X^{(1)} [\phi_\a - \xi_i(\pd u_\a/\pd x_i)]
\equiv 0$, by construction, where $X^{(1)}\id {\tt pr}^{(1)}X$ is the
first prolongation  of $X$,
so that standard Lie-point symmetries (i.e. their
generators  $Y$)  of the system (2.3) are computed by just requiring that
$$ Y^{(m)} [\De]_{S^*} \ = \ 0 \en{4} $$
where $Y^{(m)}\id {\tt pr}^{(m)}Y$ and $S^*$ is the solution manifold
for the system (2.3), and thus
corresponds to $X$-invariant solutions to $\De=0$.

Notice that these symmetries will leave globally invariant the set of
$X-$sym\-me\-tric solutions to $\De = 0$; the special symmetry $Y=X$ will
leave each of these solutions invariant. Should we look for solutions
invariant under a different vector field $X$, the system (2.3) would also
be changed; thus, conditional symmetries do not have any reason to form a
Lie algebra. We also recall that the determining equations 
for conditional \sys\ are nonlinear, see \ref{LW,Win3,Win4,Win5}.

After having so sketchily recalled the basic notions about conditional
symmetries, let us comment on the relations and differences between these
and the partial symmetries (defined above) we are discussing in this note.

Let us first comment on similarities. Comparing the definitions of
conditional and partial symmetries, it is clear that, as already remarked,
{\it any conditional
symmetry for $\De$ is also a partial symmetry for $\De$}. Indeed, if
there exists a solution $u_0 (x)$ to $\De =0$ which is $X$-invariant, we
are guaranteed of the existence of a non-empty subset $S_X$ of the solution
set $\S_\De$ which is globally $X$-invariant;  in the worst case $S_X$ 
consists of the solution $u_0$ alone: in this case we should consider
the partial symmetry to be trivial.

We also notice that the system (1.10) does obviously (by
construction) admit $X$ as a standard Lie point \sy ; this is similar to
what happens for conditional \sys , see  e.g.  \ref{Win5}. 

Let us now comment on differences between conditional and partial
symmetries.
First, we note that for partial symmetries we do not require {\it
invariance} of any solution under the vector field, but only global 
invariance of a family of \so s; indeed, here we are
looking for \so s of (1.1) satisfying also (1.14-15), but not necessarily
such that (2.2) are satisfied.

Second, in the standard conditional symmetry approach, the equation
$\De = 0$
is supplemented with a side condition [i.e. the (2.2) introduced above]
which, as just remarked, is different for different vector fields (as it
also happens for partial symmetries) but which is independent of the
differential operator $\De$ in consideration;
here instead the conditions
depend not only on the vector field $X$ but also on the equation
which we are studying and which gives origin to the hierarchy of equations
$\De^{(r)}=0$. Thus, on the one hand we aim at
identifying partial symmetries -- and through these identifying
sets of solutions --
which are more general than in the conditional symmetries approach; on
the other hand the tools we are using for this are more specific to the
single equation to be considered.

It should also be noted that in searching for ordinary Lie-point
symmetries, the determining equations for the unknown $\xi_i,\ \phi_\a$
are necessarily linear. In the conditional symmetries approach,
one supplements the differential system $\Delta=0$ with a linear equation
expressing the invariance of the solution $u(x)$ under an
undetermined vector field $X$ with  coefficients $\xi_i,\ \phi_\a$; as a
consequence of the double role of the $\phi,\xi$,  the determining
equations for the conditional symmetries, i.e. for $\phi,\xi$, are
nonlinear. For partial symmetries,
we supplement $\Delta=0$ with conditions which depend on both
$\phi,\xi$ and $\Delta$ itself, and the $\Delta^{(r)}$ could depend in
general on products of $r$ coefficients $\phi,\xi$;  if we think of
determining all the possible partial symmetries (say of given order) of
an equation $\De=0$, the determining equations for these are again and
unavoidably nonlinear.

This also means that we have no systematic way of solving or attacking
them,
and actually in general we have no hope of finding even a special
solution, i.e. a single partial symmetry, of a given equation. This
should not be a surprise, as: (i) partial symmetries are a non-generic
feature of differential equations; and (ii) conditional symmetries are a
special case of partial symmetries (see above),
and we have no algorithmic way of completely determining the
conditional symmetries of a given equation.

Thus the present method is expected
to be relevant mainly when we either have some hint -- e.g. on physical
basis -- of what partial symmetries could be, or we are specially
interested in specific candidates for partial symmetries -- again e.g.
on the basis of physical relevance -- and want to investigate if this is
the case and to determine the set $S^{(s)}$ of solutions which is globally
invariant under these partial symmetries.

Needless to say, the fact that conditional symmetries are also partial
symmetries (but the set of $X$-invariant solutions is in general only a
subset of the maximal $X$-symmetric set of solutions) provides natural
candidates for the search of nontrivial partial symmetries.

\bs\pn
{\it Remark 6.} Rather than looking for partial symmetries of a given
equation, one could be interested in the dual problem: determining all
the differential equations (say, with given dimensionality of the
dependent and independent variables and of given order) which admit a
given vector field $X$ as a partial symmetry (say, of given
order $s$). This is a hard problem, mainly because of the substitutions to 
implement in order to restrict to appropriate solution sets and manifolds: 
these make the determining equations nonlinear in $\De$. However, if we 
require $X$ to be a strong partial symmetry (see  remark 4 above) this 
problem is not present, and the equations (1.9), (1.11) are linear as 
equations for $\De$. $\odot$

\bs
Let us also stress that -- as already mentioned in the Introduction --  
our notion of $P$-symmetry can be extended 
immediately, repeating word by word the above procedure, to the case of 
generalized (or \ba ) \sys . The only difference is that
here one considers vector fields of the form  \ref{Olv}
$$X\ =\  \phi_\a(x,u,u^{(1)},u^{(2)},\ldots) \ {\pd\over{\pd u_\a}}\  
\en{5}$$
where $\phi_\a$ depend also on the derivatives 
$u_{\a,i},\ u_{\a,ij},\ldots$, with $u_{\a,i}=\pd u_\a/\pd x_i$, etc. 
(denoted globally by $u^{(1)},\ u^{(2)},\ldots$).

The case of conditional \ba\ \sys\ has been considered by \Z , 
see \ref{Zh}, where also some physically interesting examples are provided. 
But, exactly as in the case of standard Lie-point \sys , our notion of 
partial \ba\ symmetries is different from (and actually -- in some cases -- 
extends) the notion of conditional \ba\ symmetries. Example 5 in section 5
below (which is a modification of an example presented in \Z\ paper 
\ref{Zh}), although quite simple, will show in fact that a 
nonlinear PDE may possess a $P$-\ba\ symmetry $X$, and therefore may possess 
a $X$-symmetric family of \so s (i.e. a family of \so s such that the partial 
\ba\ \sy\ maps any \so\ of this family into  another of the same family, 
just  as in the case of partial Lie-point \sys ), but which are 
{\it not} invariant under this $X$: this implies that $X$ is not a 
conditional \ba\ \sy\ for the given PDE.

\section{Partial superposition principle}
\def\en#1{\eqno(3.{#1})}

We will consider here a special situation, which can be naturally
included in the above notion of $P-$symmetry.

Consider vector fields of the form
$$X=\phi_\a(x){\pd\ov{\pd u_\a}}\en{1}$$
It may be interesting to remark that, given a \df  problem $\De=0$,
applying $X^*$ to $\De$ is in this case
nothing but evaluating the Fr\'echet derivative of $\De$
applied to the vector function $\phi_\a(x)$:
$$X^*\De={\pd\De\ov{\pd u_\a}}\phi_\a(x)+
{\pd\De\ov{\pd u_{\a,i}}}\phi_{\a,i}(x)+
{\pd\De\ov{\pd u_{\a,ij}}}\phi_{\a,ij}(x)+\ldots=:\L(u,\De)\phi\en{2}$$
where  $\L(u,\De)$ is a linear
operator. If the transformation generated by (3.1) is an exact \sy , this 
implies that,
given any \so\ $u_0(x)$ of $\De=0$, then also $u=u_0(x)+\th\phi(x)$ is a
\so . But, as in the previous cases, it can happen that the \sy\ condition
$\L(u,\De)\phi\big|_{\De=0}=0$ is not satisfied in general, but only
by some subset of \so s: $\L(u,\De)\big|_{S^{(s)}}\phi=0$. This means that
$u_0(x)+\th\phi(x)$  may be a \so\ to $\De=0$ only for some special
$u_0(x)$ (and $\phi(x)$). This gives rise to a sort of ``partial
superposition principle'' for nonlinear equations. For instance, if
$$\De\id u_x+u-1+u_x^2(u_x-u_y) =0\en{3}$$
and choosing $\phi=\exp(-x-y)$, one can easily verify that
$$u^{[\th]}(x,y)=1+\th\exp(-x-y) \en{4}$$
is a $X-$symmetric family of \so s to (3.3) for any  $\la$.

The case of vector fields of the form
$$X=\phi_\a(u){\pd\ov{\pd u_\a}}\en{5}$$
is similar to the previous one (3.1), and will be considered in the next
section in the special context of dynamical systems.

\bs\pn
{\it Remark 7.} It is known that there are classes of equations admitting
{\it nonlinear } superposition principles \ref{CGM,ShW,Win0};
it has to be expected that this construction could extend to this setting,
leading to ``partial nonlinear superposition principles", but such a
discussion would go way beyond the scope of this paper. For an extension of 
the linear superposition principle, see also \ref{Wal} $\odot$

\section{Dynamical systems}
\def\en#1{\eqno(4.{#1})}
\bs
Some attention must be reserved to the special but important case of
dynamical systems, i.e. of systems of   \df\ \eq s of the form
$$\.u\ =\ f(u)\en{1}$$
where the independent variable is the time $t$ and $u_\a=u_\a(t)\in \R^n$
($p=1,\ \ell=q=n$).  Here, $\.u={\d u/\d t}$, $f=f(u)$ is assumed to be a
smooth vector-valued function (we consider for simplicity autonomous
problems, in which $f$ is independent of time).

As well known, a vector field
$$X=\phi_\a{\pd\ov{\pd u_\a}}\qq {\rm with}\qq\phi_\a=\phi_\a(u)\en{2}$$
is a Lie point (time-independent) exact \sy\ of (4.1)  if and only if
$$\[ f_\a{\pd\ov{\pd u_\a}}, \phi_\b{\pd\ov{\pd u_\b}} \]=0 \en{3}$$
which expresses just the condition $X^*\De\Big|_{\De=0}=0$, with
$\De\id (\d u / {\d t})-f(u)$.
If this commutator is not zero, the first condition $\De^{(1)}=0$, i.e.
$$\psi_\a(u)\id f_\b{\pd\phi_\a\ov{\pd u_\b}}-
\phi_\b{\pd f_\a\ov{\pd
u_\b}}\equiv(f\cdot\grad_u)\phi_\a-(\phi\cdot\grad_u)f_\a=0\en{4}$$
becomes, once $f(u)$ and $\phi(u)$ are given, a system of 
conditions for the $u_\a$, which
determines a subset (if not empty) in $\R^n$. In this situation, it may
be quite easy to verify directly if this set contains (or otherwise) a
$X-$symmetric family of \so s $u^{[\th]}(t)$.

There are some interesting and physically relevant cases in which this
situation actually occurs. Notice for instance that if one of the \so s,
say $u^{[0]}=u^{[0]}(t)$, of a $X-$symmetric family exhibits a well-defined
time-behavior
(e.g., it is periodic with some period $T$, or it is a homoclinic or
heteroclinic orbit),
and if the vector field $X$ is defined globally, i.e. along all the
time-trajectory of  $u^{[0]}(t)$, then all the \so s of the family
$u^{[\th]}(t)$ exhibit the same time-behavior (\ref{Gae}).

Given a family $u^{[\th]}=u^{[\th]}(t)$ of \so s to (4.1), and denoting by
$\L_{[\th]}(f)$ the linearization of $f(u)$ evaluated along $u^{[\th]}$,
i.e.
$$ \L_{[\th]}(f)\id \grad_u f\Big|_{u^{[\th]}}\en{5}$$
it can be useful -- in view of its applications (see below) --
to state the  above argument in the following  form:

\bigskip\pn
{\bf Proposition 2.} {\it Assume that the dynamical system (4.1) admits a
partial \sy\ $X$ of the form (4.2) and let $u^{[\th]}=u^{[\th]}(t)$ be
an orbit of \so s obtained under the
action of the group generated by  $X$. Then $\phi=\phi(u^{[\th]})$
satisfies the equations
$$\.\phi=\L_{[\th]}(f)\cdot\phi\en{6}$$
and 
$$\phi={\d u^{[\th]}\ov{\d\th}}\ .\en{7}$$ }
\bs

In order to prove this, just note that
$\.\phi=\.u\cdot\grad_u\phi=f\cdot\grad_u\phi$,
then (4.6-7) come from (4.4) and from equations (1.14-15).

\bs\pn
{\it Remark 8.}
This Proposition is relevant, e.g., in the cases where the dynamical
system admits a manifold of homoclinic (or heteroclinic) orbits.
Indeed Proposition 2 ensures that the
vector $\phi={\d u^{[\th]} /{\d\th}}$, tangent to the family
$u^{[\th]}$, is a bounded \so\ (for all $t\in\R$) of the eq. (4.6),
which is usually called the ``variational \eq '', obtained by linearizing
the dynamical system along  $u^{[\th]}$.
On the other hand, the knowledge of all  bounded \so s to the
variational  \eq\ is important to construct the Mel'nikov vector, which
provides a useful tool for determining the onset of chaotic behavior in the
dynamical system in the presence of perturbations. For the
applications of this fact to the theory of chaotic behaviour of
dynamical system, which
clearly goes beyond the scope of this paper, we refer, e.g., to
\ref{ch1,ch2,ch3}.
$\odot$

\bs\pn
{\it Remark 9.}
In the same context as in Remark 8, it is well known that
another bounded \so\ of the variational \eq\ (4.6) is provided
by  the time-derivative $\d u/\d t$; it can be remarked that, from the
group-theoretical point of view, the two tangent vectors $\d u/\d\th$ and
$\d u/\d t $ are to be considered perfectly on the same level,  indeed the
time evolution of the dynamical system is always a (non-linear) \sy\
for the system, with generator
$$  f_\a{\pd\ov{\pd u_\a}\ }=\ {\d\ov{\d t}}\en{8}$$
where the time $t$ plays the role of the parameter $\la$. $\odot$
\bs\pn

\section{Examples I: PDE's}
\def\en#1{\eqno(5.{#1})}

\bs
As pointed out in the above section 1, our procedure may be fruitful if
one is able to conjecture some ``reasonable  candidate'' for such a
partial \sy . As said before, one may consider first of all the
conditional \sys : examples 1 and 4 will cover situations where in fact
one finds $X-$symmetric sets of \so s which contain as a special case a
$X-$invariant \so\ (which could be also obtained via the standard method
of conditional \sys ). Examples 2 and 3, instead, will show cases where
the partial \sy\ is {\it not} a conditional \sy , and in fact we are
able, introducing suitable $P-$symmetries, to
obtain $X-$symmetric sets of \so s which do not contain any $X-$invariant
\so . The same applies to example 5, dealing with generalized symmetries.

Another typical situation which may suggest possible candidates as
$P$-symmetries occurs for instance, as illustrated by the
examples 1 and 3 below, when the \df  problem is written as a sum
of two terms, the first one possessing a known group of exact \sys , plus a
``perturbation'' term which breaks these \sys . Then the natural candidates
are just the \sys\ of the unperturbed term.

Other  convenient situations may occasionally
occur  when imposing the chain of conditions
$(X^*)^r\De^{(0)}=0$ leads to a simpler \df problem (e.g. thanks to the
vanishing of some term in the \eq , as in the examples 1,2,3 below).
Also, observing  that in our procedure each subset $S^{(r)}$ is in general
considerably smaller than the preceding sets in
the chain $S^{(0)}\ss\ldots\ss S^{(r)}$,
it can happen that, after just one or very few steps, one is able to
``isolate'' directly ``by inspection'' within some set $S^{(r)}$ (although
$X^*\De^{(r)}\Big|_{S^{(r)}}\not=0$) some $X-$symmetric  family of \so s
(see example 1).
\bs\pn
{\bf Example 1}  (Modified Laplace equation). Consider the PDE, putting
$x_1=x,\ x_2=~y$, and with $\ell=1,\ q=1,\ m=3$,
$$\De\id u_{xx}+u_{yy}+g(u)u_{xxx}=0\en{1}$$
and the vector field, generating the rotations in the plane $x,y$,
$$X=y{\pd\over{\pd x}}-x{\pd\over{\pd y}}\en{2}$$
which is not a \sy\ for (5.1) unless $g(u)=0$ (we shall show below that
it is {\it not} enough to impose $u_{xxx}=0$).
The first application of the prolonged \sy\ $X^*$ to (5.1) gives
$$\De^{(1)}\id X^*\De\Big|_{\De=0}=3g(u)u_{xxy}\en{3}$$
and then, excluding constant \so s $u_0$ satisfying $g(u_0)=0$,
we get  the first condition
$$u_{xxy}=0\en{4}$$
Applying the convenient prolongation $X^*$ to this equation  gives
$$\De^{(2)}=2u_{xyy}-u_{xxx}\en{5}$$
which does {\it not} satisfy $\De^{(2)}\Big|_{S^{(1)}}=0$, and therefore
one obtains the other condition $2u_{xyy}-u_{xxx}=0$.
Iterating the procedure, one has that
$$\De^{(3)}=X^*\De^{(2)}=-7u_{xxy}+2u_{yyy}\en{6}$$
which again is not zero on the set
$S^{(2)}$, but -- using now (5.4) --  gives the new condition
$$u_{yyy}=0\en{7}$$
Then, other steps are necessary; proceeding further, we get
$X^*u_{yyy}=-3u_{xyy}$, giving $u_{xyy}=0$
and, finally, from another application of $X^*$ to this, we get
$$X^*u_{xyy}=-2u_{xxy}+u_{yyy}=0\en{8}$$
which is in fact zero, thanks   to (5.4) and to (5.7).

Then, in conclusion, (5.2) is a $P$-symmetry of order $s=5$ for 
the equation (5.1). 

The set $S^{(s)}$ of the simultaneous \so s of all the above
conditions has the general form
$$S^{(s)}=\big\{u(x,y)=A(x^2-y^2)+Bxy+Cx+Dy+E\big\}\en{9}$$
where $A,\ldots,E$ are arbitrary constants, and it is easy to recognize,
putting, e.g., $A=a\cos 2\th,\ B=a\sin 2\th$ and respectively
$C=c\cos\th,\ D=c\sin\th$,
that this set contains, apart from  the constant \so s\ $u(x,y) = E$, 
which are clearly rotationally invariant, two different
families of orbits of \so s to the initial problem (5.1), which in fact
are transformed into themselves under rotations.

It is easy to see that the rotation \sy\ $X$ is not only a $P-$symmetry
but also a conditional \sy\ for the problem (5.1), indeed, the
rotationally invariant \so\ must be of the form  $u=v(\rho)$ with
$\rho=(x^2+y^2)/2$; substituting into (5.1), we get
$$2\rho v''+v'+g(v)(3xv''+x^3v''')=0\en{10}$$
which (for $g(v)\not= 0$) implies  $v'=0$. Thus the only
rotationally invariant \so s are in this case trivially given by
the constant ones, which are in fact included in the  larger
$X-$symmetric set of the \so s (5.9) found above.

The result (5.9) looks quite
obvious and indeed could be expected just after one step (or perhaps
immediately); this example however can be useful for several reasons.
First of all, it shows that it could be possible to reach the conclusion by
means of an iterative procedure. Second, it also shows that it is {\it not}
sufficient to impose, together with (5.1), only the condition of the
vanishing of the ``symmetry-breaking'' term
$$u_{xxx}=0\en{11}$$
indeed this equation does {\it not}
admit the rotation \sy , therefore a simultaneous \so\ of both (5.1) and
(5.11), e.g. $u=x^2y-y^3/3$, would be transformed by $X$ into a \so\ of
$u_{xx}+u_{yy}=0$ but {\it neither} of (5.1) {\it nor} of (5.11).
Similarly, it is not sufficient to impose that the \so s of the initial
equation (5.1) satisfy only the first condition (5.4). E.g., with
$g(u)=1$, the \so\ $u(x,y)=\exp x$ of (5.1) satisfies also (5.4)
but does not belong to any family of \so s of (5.1) which is also
globally invariant under rotations. \hfill $\triangle$

\def\D{\De}
\def\pa{\pd}
\def\tD{\~\D}
\bs\pn
{\bf Example 2} (KdV equation).
Consider, with $x_1=x$, $x_2=t$, the classical Korteweg-de Vries equation
$$\De\id u_t+u_{xxx}+uu_x=0\ .\en{12}$$
It is well known that it admits an exact scaling symmetry, given by
$$  X \ = \ -2\ u {\pd\ov{\pd u}} + \ x {\pd\ov{\pd x}}+3\ t{\pd\ov{\pd t}}
   . \en{13}$$
We now want to determine if there are scaling $P$-symmetries for the
KdV.

We consider the generic scaling vector fields
$$ X \ = \ a\ u {\pd\ov{\pd u}} + b\ x {\pd\ov{\pd x}}+c\ t{\pd\ov{\pd
t}}\en{14}$$
(notice that $X=bX_0$, where $X_0$ is the exact \sy\ (5.13), for $a=-2b$,
$c=3b$). Applying the (third) prolongation $X^*$ of $X$ on $\D$, we obtain
$$\begin{array}{rl}
X^* \D\id \D^{(1)} \ = &\ a \D + [-c u_t + a u u_x - b u u_x
- 3 b u_{xxx}] \\ 
   = &\ (a-c) \D + [ (a-b+c) u u_x - (3b-c) u_{xxx} \ ; \end{array}
\en{15}$$
requiring to be on $S^{(0)}$, i.e. on $\D = 0$
(the solution manifold of the KdV), i.e.
performing the substitution
$ u_t \to - (u_{xxx}  + u u_x )$,
we obtain the condition
$$ \~\D_1\id\De^{(1)}\Big|_{S^{(0)}}
   = \ (a-b+c) u u_x - (3 b - c) u_{xxx}=0 \ ; \en{16}$$
notice this is identically zero only in the ``trivial'' case $X = b X_0$ 
where $X_0$ is the exact \sy\ (5.13). We rewrite this as
$$ \~\D _1 = \ A u u_x - B u_{xxx}=0 \ , \en{17}$$
and consider different cases.

{\it Case I.} If $A=B=0$ we are, as already remarked, in the case
$X=bX_0$
and we are thus considering the case of exact scaling symmetry.

{\it Case II.} If $A\not=0$ and $B=0$, $\~\D _1 =0$ reduces to $u u_x
=0$,
which in turn implies $u(x,t) = \a (t)$ and, due to $\D=0$, $u(x,t) =
c_0$. We reduce then to the trivial case of constant solutions (these are
obviously
transformed among themselves under the action of any $X$ of the form
(5.14), and
are invariant under any field with $a =0$ for the solution with $c_0 \not= 0$, 
and under any $X$ when $c_0 =0$).

{\it Case III.} If $A=0$ and $B \not=0$, then $\~\D _1 =0$ reduces to
$u_{xxx} =0$; notice that at next step we have $X^* (\~\D _1) =
(a-3b) u_{xxx}$ which is obviously zero on $S^{(1)}$.
The solution set for $u_{xxx}=0$ corresponds to $u(x,t) = \a (t) +
\b (t) x + \gamma (t) x^2$; substituting this into the KdV equation
we obtain that $\gamma (t) = 0$ and that
$$ \cases{ \a ' + \a \b = 0 & \cr \b' + \b^2 = 0 &. \cr} \en{18}$$
The second of these yields $\b (t) = (c_1 + t)^{-1}$ and using this we
also get $\a (t) = c_2 (c_1 + t)^{-1}$.
Thus, $X$ is a $P-$symmetry and the set of solutions to the KdV which is
globally invariant under $X$, with $A=0$ and $B\not=0$, is given by
$$ u(x,t) \ = \ {c_2 + x \over c_1 + t} \en{19}$$
with $c_1 , c_2$ arbitrary constants.
Note that no solution of this form is invariant under such $X$: thus,
these $P-$symmetries $X$ do {\it not} correspond to conditional symmetries.
As two different examples of this case one can consider in (5.14) 
$a=0,\ b=c=1$, which generates the simultaneous dilations of the independent 
variables $x$ and $t$ (and then, in terms of the Lie parameter $\lambda$, 
under the action of $X$ (cf. (1.14) and (1.15)), one has in (5.19)
$c_2=\exp\lambda, \ c_1=c\exp\lambda$, with arbitrary $c$);
or  $b=0,\ a=-1,\ c=1$, which generates the simultaneous dilation of the
independent variable $t$ and shrinking of the dependent variable $u$
(and then $c_1=\exp\lambda, \ c_2=c$).

{\it Case IV.} If $A\not=0$ and $B\not=0$, we have a more interesting
case; now $\~\D _1 =0$ reads
$$ u_{xxx} = {a-b+c \over 3b-c} u u_x \ . \en{20}$$
Applying $X^*$ on $\D^{(1)}$, we obtain
$$\begin{array}{rl}
\D^{(2)} \ := &\
(a-c)^2 \D + (a-c) \tD_1 + A(2a-b) u u_x - B (a-3b) u_{xxx} \\ 
= & \ (a-c)^2 \D + (3a - b - c) \tD_1 + B (a + 2b ) u_{xxx} \  ; \\ 
\end{array} \en{21}$$
when we impose (5.16) and (5.20), this reduces to
$$\~\D_2\id\De^{(2)}\Big|_{S^{(1)}}  \ =
\ (a + 2 b) \ u u_x \ = \ 0 \ . \en{22}$$
If $a+2b \not=0$, this requires $u_x =0$,
i.e. we are reduced to the case of trivial constant solutions.
On the other hand, if
$$ a + 2b \ = \ 0\ , \en{23}$$
which also implies $A=B$, we have obtained that  $X$ is a $P$-symmetry
of the KdV.

It should be noted that now, thanks to $A=B$, the equation $\tD_1 =0$
reads $u_{xxx} = u u_x$ for all $X$ in this class; this equation can be
reduced to ``quadratures'', and
for functions $u(x,t)$ satisfying this, the KdV reduces  to
$$u_t=- 2 u_{xxx} = -2 u u_x=-2u\big[{u^3/3}+\a(t)u+\b(t)\big]^{1/2}\ 
.\en{24}$$
Notice also that in this case we have $X = a [u (\pa / {\pa u}) -
2x (\pa / {\pa x}) ] + c (\pa / {\pa t})$;
considering the first two components of $X$ is sufficient to guarantee
that -- as can be seen by solving the characteristic equation for $X$ --
the only solution invariant under any such $X$ is the trivial one
$u(x,t) = 0 $.

This completes the possible cases in the analysis of (5.14).
\hfill$\triangle$
\bs\pn
{\bf Example 3} (A nonlinear heat equation).
Consider this nonlinear heat equation
$$\De\id u_t-u_{xx}-uu_{xx}+u_x^2=0\en{25}$$
and the vector field
$$X=2t{\pd\ov{\pd x}}-xu{\pd\ov{\pd u}}\en{26}$$
(which is an exact \sy\ of the standard linear heat \eq ). One gets at the
first step
$$\De^{(1)}=X^*\De=x(u_x^2-uu_{xx})\en{27}$$
In this case the two conditions
$$\De=\De^{(1)}=0\en{28}$$
are enough to define a set $S^{(1)}$ of \so s which is $X-$symmetric,
indeed, one gets
$$X^*\De^{(1)}\Big|_{S^{(1)}}=0  \en{29}$$
and therefore (5.26) is a $P$-symmetry of order $s=1$ for the equation 
(5.25). The \eq s (5.28) can be easily solved to get the $X-$symmetric 
family of \so s
$$u^{[\th]}(x,t)=c\exp(-x\th+t\th^2)\en{30}$$
where $c$ is a constant, which is indeed transformed into itself by
the finite transformations generated by $X$, i.e.
$$t\to t'=t, \qq x\to x'= x+2t\th, \en{31}$$
$$u\to u^{[\th]}=u(x,t)\exp(-x\th-t\th^2)=
u(x'-2t'\th,t')\exp(-x'\th+t'\th^2)\ .\en{32}$$
We can also verify that the above \tr\ (5.26) is not a (nontrivial) 
conditional \sy\ for
the problem (5.25). Indeed, the functions $v=v(x,t)$ satisfying the
invariance condition (2.3) must be of the form
$$v=w(t)\exp(-x^2/4t)\en{33}$$
Inserting in (5.25) gives
$$-2t\ {w'\ov w}\ =\ 1+w\exp(-x^2/4t)\en{34}$$
which can be satisfied only by $w\equiv 0$. This agrees with our previous
result (5.30), which shows in fact that no \so s of the form (5.33) is
included in the family (5.30). \hfill $\triangle$

\bs\pn
{\bf Example 4} (The Boussinesq equation.) The  Boussinesq equation
$$\De\id \ u_{tt} + u u_{xx} + (u_x)^2 + u_{xxxx} = 0\en{35}$$
has been used as a testing ground for conditional symmetries
\ref{ClK,LW,Win4,Win5}, and thus is appropriate to (briefly)
consider it from the point of view of partial symmetries as well.
We consider here only the first (and simplest) one  of the conditional
\sys\ of this \eq\ \ref{LW,Win4,Win5}, namely
$$X = {\pd\ov{\pd t}} + t {\pd\ov{\pd x}} - 2 t {\pd\ov{\pd u}} \en{36}$$
Applying our procedure, we find
$$\De^{(1)}\id X^*\De\ = \ u_{xt} + t u_{xx}
\qq\qq\De^{(2)}\id X^* \De^{(1_)}\equiv 0\en{37}$$
We then have to look for the simultaneous \so s of the two \eq s $\De=0$
and $\De^{(1)}=0$. The set $S^{(1)}$ of these $X-$symmetric
\so s is not empty, in fact
it must contain at least the $X-$invariant \so s to (5.35), which can be
obtained via the conditional \sys\ approach \ref{LW,Win4,Win5};
actually, we shall see that, as in Example 1, this set is much larger.
Solving the first condition $\De^{(1)}=0$ gives indeed
$$u(x,t)=w(x-t^2/2)+g(t)\en{38}$$
where $w$ and $g$ are arbitrary; notice that only if $g=-t^2$ this \so\
is invariant under $X$. We then put for convenience
$$g(t)=-t^2+h(t)\en{39}$$
Inserting this into the Boussinesq \eq , we find that $w$ and $h$ must
satisfy
$${\d\ov{\d z}}(w''' + w w' - w-2z) + hw''+{\d^2 h\ov{\d t^2}}=0\en{40}$$
where $z=x-t^2/2$ and $w=w(z)$.
Now, if $h=0$, the well known \eq\  for $w(z)$ \ref{LW,Win4,Win5} is
recovered, but with
$h\not=0$ other \so s of the Boussinesq \eq , not invariant  under 
(5.36), can be found. For instance, we obtain the following family of \so s
$$u(x,t)= w(z)+A\en{41}$$
where $A$ is a constant and $w=w_A(z)$ satisfies the equation
$$  w''' +  w w' -  w+A w'=c +2z\en{42}$$
and also the other family of \so s (quite trivial, but not included in the
previous set (5.41))
$$u(x,t)= Bx-{B^2\ov 2}t^2 +Ct+D\en{43}$$
with $B,C,D$ constants. It is clear that all the $X-$invariant \so s found
via the conditional \sy\ approach are recovered for particular values of
the
parameters $A,B,C,D$. \hfill $\triangle$

\bs\pn
{\bf Example 5}
(A PDE admitting a $P$-\ba\ symmetry). Finally, we deal with an example 
of the extension of partial \sys\ to \ba\ symmetries, mentioned at the 
end of section 2. 
Consider, as in example 4 of \Z\ paper \ref{Zh}, a PDE of the form
$$u_t\ =\ u_{xx}\ +\ R(u,u_x) \en{44}$$
and the \ba\ vector field
$$X\ =\ (u_{xx}-a\ u)\ {\pd\over {\pd u}}\qquad\qquad (a\in \R) \en{45}$$
According to the prescriptions for the conditional \sys , \Z\ looks for \so s 
of (5.44) restricted to the manifold of the {\it invariant} \so s under 
the transformations generated by the vector field (5.45), i.e. of the \so s 
satisfying also
$$u_{xx}\ -\ a\ u\ =\ 0\en{46}$$
(and its differential consequences), and concludes that (5.45) is a 
conditional \ba\ \sy\ for (5.44) if and only if the nonlinear term $R$ 
in (5.44) has a special form, see \ref{Zh}.

We choose instead, as an example, 
$$R\ =\ u_x^2\ -\ {a\over{2}}\ u^2\en{47}$$
which has {\it not} the above form, and -- obviously -- we do {\it not} 
impose invariance under $X$. Applying the prolongation $X^*$ to the PDE 
(5.44) with a generic $R$, one finds
$$X^*\De\ =\ -aR+auR_u+au_xR_{u_x}+R_{uu}u_x^2+2R_{uu_x}u_xu_{xx}+
R_{u_xu_x}u_{xx}^2 \en{48}$$
where $R_u=\pd R/\pd u$, etc.;  with our choice (5.47), this gives
$$X^*\De\ =\ 2\Big(u^2_{xx}-{a^2\over {4}}u^2\Big) \en{49}$$
According to our procedure, we look for the \so s of
the equation $\De^{(1)}\id X^*\De=0$, which are given by (let $a>0$)
$$u_+(t,x)=\phi_+(t)\exp\big(\sqrt{a/2}\ x\big) \qquad {\rm or}\qquad
  u_-(t,x)=\phi_-(t)\exp\big(-\sqrt{a/2}\ x\big) \en{50}$$
(and which clearly -- as expected -- do {\it not} satisfy the $X$-invariance 
condition (5.46)). It is now easy to see that our other condition
$\De^{(2)}\id X^*\De^{(1)}=0$  is satisfied once restricted to the  
\so s (5.50), showing that these constitute a $X$-symmetric set, 
although not $X$-invariant. Finally, taking into account the equation
$\De^{(0)}=0$, i.e. the equation (5.44), one finds the two families of 
\so s $u_+(t,x)$ and $u_-(t,x)$ of the PDE (5.44)
$$u_{\pm}(t,x)\ =\ c_{\pm} \exp\Big((a/2)\ t\ \pm\ \sqrt{a/2}\ x\Big)\en{51}$$
(notice incidentally that no combination of $u_+(t,x)$ with $u_-(t,x)$ solves 
the PDE). It is immediate to see that each one of these families 
is mapped into itself by the vector field $X$, and therefore we can 
conclude that $X$ is a $P$-\ba\ symmetry (but not a conditional \ba\ 
\sy ) for the PDE (5.44) with the nonlinear term given in (5.47).
\hfill $\triangle$

\section{Examples II: dynamical systems}
\def\en#1{\eqno(6.{#1})}

In this section we briefly consider some cases where the discussion
of the above section 4 can be applied.
\bs\pn
{\bf  Example 6.} Consider a three-dimensional dynamical system, with $u 
\in \R^3 $, $u \id  (x,y,z) $, of the form
$$\begin{array}{rl}
\.x=&x(1-r^2)-y+zg_1(x,y,z)\\ 
             \.y=&y(1-r^2)+x+zg_2(x,y,z)\\ 
             \.z=&zg_3(x,y,z)
\end{array} \en{1}$$
where $g_\a(x,y,z),\ \a=1,2,3$ are arbitrary smooth functions and
$r^2=x^2+y^2$.
It is easy to verify that considering the vector field, generating
rotations in the plane $(x,y)$,
$$X=y{\pd\over{\pd x}}-x{\pd\over{\pd y}}\en{2}$$
the partial symmetry condition (4.3) takes the form
$$zG_\a(x,y,z)=0\qq\qq \a=1,2,3\en{3}$$
where, for instance,
$G_1(x,y,z)=g_2-y(\pd g_1/\pd x)+x(\pd g_1/\pd y)$, which is nonzero 
for generic $g_\a$. As obvious in
this simple example, the dynamical system exhibits rotation \sy\ once
restricted to the plane $z=0$, and in this plane one can find three
different
families of \so s $u^{[\th]}(t)$ which are mapped into themselves by
the rotations: the trajectories lying in $r^2<1$, and resp. in
$r^2>1$, spiralling
towards the limit cycle $r^2=1$, and the \so s running on the single 
trajectory which is left fixed by the partial \sy\ (the limit cycle).
\hfill $\triangle$
\bs\pn
{\bf Example 7.} This example is, admittedly, a somewhat artificial
one. Indeed, it
has been constructed to put together, in a non-symmetric dynamical system,
the presence of a partial nonlinear \sy , and of a two dimensional
heteroclinic manifold. Let us consider then, with
$u\id(x,y,z)\in\R^3$, the system
$$\begin{array}{rl}
\.x=&x\big(1-z\exp(-y)\big)+g_1(x,y,z)(R^2-z)^2\\ 
             \.y=&y\big(1-z\exp(-y)\big)+g_2(x,y,z)(R^2-z)^2\\ 
             \.z=&-z+yz\big(1-z\exp(-y)\big)-z^2\exp(-y)+3R^2
\end{array} \en{4}$$
where $g_1,\ g_2$ are arbitrary smooth functions, and $R^2=
{1\ov 6}(x^2+y^2)\exp(+y)+{1\ov 2} z^2\exp(-y)$. Notice first of all that,
if $g_1=g_2=0$, the dynamical system would admit the (exact) nonlinear \sy
$$X=y{\pd\over{\pd x}}-x{\pd\over{\pd y}}-xz{\pd\ov{\pd z}}\en{5}$$
This \sy\ has been introduced in \ref{CG}, where it has been also shown
that the most general dynamical system admitting this \sy\ has the form
$$\begin{array}{rl}
\.x=&xf(r^2,v)+yg(r^2,v)\\ 
             \.y=&yf(r^2,v)+yg(r^2,v)\\ 
             \.z=&zh(r^2,v)+yzf(r^2,v)+xzg(r^2,v)
\end{array} \en{6}$$
where $f,g,h$ are
arbitrary functions of the quantities $r^2=x^2+y^2$ and $v=z\exp(-y)$.
It is not difficult to show that the dynamical system (6.4) possesses a
two-dimensional
manifold $u^{[\th]}$ of heteroclinic orbits, joining biasymptotically
the critical points $O=(0,0,0)$ and $A=(0,0,2)$, and given by
$$\begin{array}{rl}
u^{[\th]}(t,t_0,\th)\equiv\big(
\sqrt{3}&{\rm sech }(t-t_0) {\cos }\th, \
\sqrt{3}\ {\rm sech}(t-t_0) {\sin }\th,\\ 
         &\big(1+{\rm tanh}(t-t_0)\big) {\rm exp}\big(\sqrt{3}\
         {\rm sech}(t-t_0){\sin }\th\big)\big)
\end{array} \en{7}$$
where $t_0$ is arbitrary. Although the dynamical system (6.4) does not
admit in general the
\sy\ (6.5), we can easily check that this is a partial \sy\ for the 
system (6.4), and in fact each one of the heteroclinic
orbits in the manifold (6.7), obtained  keeping fixed $\th$ and varying $t$,
is transformed into another orbit of the same manifold by the
transformations generated by  (6.5).
Indeed, the finite action of this \tr\ on the coordinates is given by
$$\begin{array}{rl}
x&\to x'=x\cos\th+y\sin\th\\ 
    x&\to y'=-x\sin\th+y\cos\th\\ 
    z&\to z'=z\exp(-x\sin\th+y\cos\th)=z\exp(y')
\end{array} \en{8}$$
According to the remarks 8 and 9 of Sect. 4, we can also directly
verify that the two tangent vectors $\d u/\d t$ and $\d u/\d\th$
are \so s of the variational equation (4.6) obtained from (6.4).
\hfill $\triangle$


\section*{Appendix. Discrete partial symmetries}
\def\en#1{\eqno(A.{#1})}

It should be noted that the construction and results proposed here, and 
discussed in the framework of continuous Lie-point transformations, do
also apply to more general kind of transformations, such as non Lie-point 
ones (see \ref{Olv}; we have briefly considere here the case of B\"acklund 
symmetries) and  discrete 
Lie-point  transformations. In this appendix we briefly discuss the 
application of our approach to the latter case.

In this respect, we would like to recall that the main
obstacle for the use of discrete symmetries in connection with
differential equations is the difficulty in their determination: indeed,
except for discrete symmetries which are immediately evident (such as
parity transformation or shift by a period) we have no algorithmic way
for solving the determining equations for discrete symmetries; this is
due to the fact in this case we cannot reduce to the tangent space of
suitable manifolds, and thus the determining equations are nonlinear. In
the present case, nonlinearity is already present for continuous
$P$-symmetries, and thus determination of possible discrete
$P$-symmetries is a comparably difficult task; as already mentioned in
discussing continuous ones, we have some hope of success only if we are
led by physical considerations or if we want to analyze (again on
physical basis) specific kind of transformations. Notice however that in
this respect there are several discrete transformations to be
considered, which are natural in physical terms and which are quite
interesting if happening to be $P$-symmetries: these are reflections and
discrete translations. In some contexts, e.g. in systems relevant in
statistical mechanics \ref{Car}, one would also be specially interested in
discrete scale transformations.

The similarity between the study and determination of discrete and
continuous $P$-symmetries is particularly transparent in terms of the
previous remark~3.

We can thus consider a general map $R : (x,u) \to (\~x , \~u )$ and its
prolongation $R^*$ acting on $(x,u^{(m)} )$; we apply this on the
differential equation $\Delta$. If
$$ R^*  \Delta \Big|_{\De=0} = 0 \en{1} $$
then $R$ is a discrete Lie-point exact symmetry of $\Delta=0$;
we assume
(A.1) is not satisfied, and write
$$ \Delta^{(1)}\id  \  R^* (\Delta ) \ . \en{2} $$
We will then consider the common solution set $S^{(1)}$ of $\Delta$ and
of $\Delta^{(1)}$, and consider $R^* (\Delta^{(1)})$ on this; if this is
nonzero, we will iterate the procedure as in the continuous case, until we 
reach an $s$ such that $R^* \Delta^{(s)} \big|_{S^{(s)}} = 0 $. This 
$S^{(s)}$ identifies a set of solutions to $\Delta =0$ which is 
$R$-symmetric, and our results apply in this setting as well.
Notice however that in this case we cannot iterate our procedure 
indefinitely if, as it happens for many interesting discrete
transformations, there is a $k>0$ such that $R^k = I$.

\bs\pn
{\it Example A1.} Consider the equation
$$ \Delta\id \ u_{xx} + u_{yy} + g(u) u_{xxx} \ = \ 0 \en{3} $$
and the discrete transformation corresponding to $x$-reflection,
$$ R : (x,y;u) \to (-x , y , u) \ . \en{4} $$
It is easy to see that $R^*$ leaves $g(u)$, $u_{xx}$ and $u_{yy}$
invariant, and maps $u_{xxx}$ in minus itself; thus,
$$ \Delta^{(1)} \id  R^*  \Delta  \ = \ u_{xx} + u_{yy} - g(u)
u_{xxx} \en{5} $$
which on $S^{(0)}$ yields
$$ \~\Delta^{(1)} \ = \ - 2 \,  g(u) \, u_{xxx} \ . \en{6} $$
Therefore, $S^{(1)}$ corresponds to solutions of $\Delta=0$ satisfying
the
additional condition $u_{xxx}~=~0$; notice that with this $\Delta=0$
reduces to the wave equation $u_{xx} + u_{yy} = 0$ restricted to the
space of functions $u(x,y) = \alpha (y) + \beta (y) x + \gamma (y) x^2$.
Therefore, we have $\beta'' (y) = \gamma'' (y) = 0$, and $\alpha '' (y)
= -2 \gamma (y) $ (which in turns implies $\alpha^{iv} (y) = 0$).
   \hfill $\triangle$
\bs\pn
{\it Example A2.} Let us consider a system with boundary conditions
$u(0,t) = u (2 \pi , t) = 0$ and depending on an external constant
$\mu$, i.e.
$$ \Delta\id \ u_t - \mu u - u_{xx} + u_{xxx} \ = \ 0 \ .
\en{7} $$
It is easy to see, passing in Fourier representation, that the solution 
$u_0 (x,t) \equiv 0$ is stable
for $\mu < 1$, while for $\mu > 1$ this is unstable and we have
instead stable periodic solutions; notice that looking for solutions in
the form $u (x,t) = f_{k \omega } \exp [i (k x + \omega t)$ the
dispersion relations result to be
$$ \omega = - i (\mu - k^2 ) \ , \en{8} $$
and the boundary conditions impose $k$
is an integer. The consideration of higher order terms would permit to
obtain $u(x,t)$ as a Fourier series in terms of functions $x$-periodic
of period $2 \pi$ and higher harmonics, i.e.
$$ u(x,t) = \ \sum_{k=1}^\infty f_k (t) \sin (k x) \ . \en{9} $$

We will consider the discrete transformation corresponding to shift of
$\pi$ in $x$, i.e.
$$ R : (x,t;u) \to (x + \pi , t ; u) \ ; \en{10} $$
notice this does not in general respect the boundary conditions.

It is easy to see that $R^* \Delta\big|_{\De=0} = \mu [u (x,t) \, -
\, u (x - \pi , t) ] $ and thus reduction to $S^{(1)}$ correspond to
$$ u (x,t) \, - \, u (x - \pi , t) \ = \ 0 \ , \en{11} $$ i.e. to
the requirement that only even harmonics are present in the Fourier
expansion for $u(x,t)$, i.e. in (A.8) all the $f_k (t)$ for odd $k$ are
identically zero. Notice this means in particular that the fundamental
wave number for $u(x,t)$ will not be 1, but 2.

We remark, although this goes beyond the limits of the present paper,
that when $\mu$ is not a constant but a varying external control
parameter, the problem (A.7) presents a Hopf bifurcation at $\mu =
1$; if we restrict to the subset of solutions $S^{(1)}$, i.e. if we
impose the additional boundary condition (A.10), we still have a Hopf
bifurcation, but now at $\mu = 4$.  \hfill $\triangle$

\vfill\eject

\end{document}